\documentclass[preprint,12pt]{elsarticle}
\usepackage{amsfonts, amssymb, amsmath, mathrsfs}
\usepackage[top=1in,bottom=1in,left=1.25in,right=1.25in]{geometry}
\usepackage{xypic}

\begin{document}
\newtheorem{lemma}{Lemma}[section]
\newtheorem{proposition}{Proposition}[section]
\newtheorem{theorem}{Theorem}[section]
\newtheorem{corollary}{Corollary}[section]
\newtheorem{example}{Example}[section]
\newtheorem{definition}{Definition}[section]
\newtheorem{remark}{Remark}[section]
\newtheorem{property}{Property}[section]
 \makeatletter
    \newcommand{\rmnum}[1]{\romannumeral #1}
    \newcommand{\Rmnum}[1]{\expandafter\@slowromancap\romannumeral #1@}
    \makeatother

\begin{frontmatter}
\title{A study on central soft sets: Definitions and basic operations}
 \author[label1]{Guan Xuechong\corref{cor1}}
 \ead{guanxc@foxmail.com}
 \cortext[cor1]{Corresponding author at: College of Mathematics and Statistics, Jiangsu Normal University,
China.}
\address[label1]{College of Mathematics and Statistics, Jiangsu Normal University, Xuzhou, 221116, China}
\begin{abstract}

In this paper, a new kind of soft sets related with some common decision making problems in real life called
central soft sets is introduced.
Properties of some basic operations on central soft sets are shown.
It is investigated that some classic operations between soft sets can be obtained
by central soft sets with selecting different central sets.
We initiate the concepts of an evaluation system for a parameters set and its optional solutions.
An algorithm is presented to solve such decision making problems.

\end{abstract}


\end{frontmatter}

\section{Introduction }

In order to describe some uncertainties which are appeared around
everywhere, Molodtsov \cite{first,optimization,soft} innovated a
novel concept of soft sets as a new mathematical tool for solving
these problems. A soft set, in fact, is a tuple which associates
with a set of parameters and a mapping from the parameter set into
the power set of an universe set. It is a parameterized family of
subsets of the universe set. At present soft set theory has combined
with several directions such as universal algebra
\cite{rings,group,semirings, algebra}, relation analysis
\cite{operations,operation,continuous,ideal,Feng,Aktas,Babitha} and
other mathematical domains \cite{fuzzy,topology,category,logic}.
Especially this theory has been applied in decision-making problems
\cite{reduction, decision,Deli,adjustable}.

Soft set theory has its convenience and simplicity in application including decision-making.
We try to solve the following type of problems by soft sets.
For example, this case is often encountered in our real life, that a jury composed of more than one person will
make investigations on a project.
The contents of this project needed to be examined are involved with multiple fields
which have some certain sufficient knowledge background.
Each member of the jury will make an appropriate judgment on the basis of what they know.
However, everyone has a specialized area in which they excel generally.
So naturally, their relevant scoring in non specialized field need for discretion.
In the end we need to make a comprehensive evaluation based on these inspection results.
Concerned about this phenomenon, the concept of central soft sets is proposed in this paper.
It is different from the study which focuses on valuation objects of attribute sets \cite{fuzzy,interval,int,Prob},
that we pay attention to central attribute sets of soft sets, and examine
what role they will play in the operations between soft sets.
In the part of theoretical application, we define evaluation systems of a parameters set consist of central soft sets.
Then an algorithm for making corresponding decision of an evaluation system is given.

The rest of this paper is organized as follows.
The second section briefly reviews some basic notions on soft sets firstly.
The concept of central soft sets is proposed.
The properties of some basic operations on central soft sets over a universe set are given in detail.
We research the relationship between operations defined in this paper and some classic operations in soft set theory.
In the last section, we will study evaluation systems for parameters sets and give the method of obtaining solution of them.

\section{Preliminaries}

First we present some basic definitions and notations used in what follows. In this paper,
$U$ is an initial universe set. The symbol ${\cal P}(U)$ denotes the power set of $U$.
Let $E$ be a set of parameters which usually are initial attributes, characteristics, or properties of objects in $U$.

To try to solve problems by an intuitive, simple and practical way is an important and distinguishing feature
in the study of soft sets.
This is also a reason why we introduce the concept of central soft sets here.

\begin{definition}\label{definition:1}
{\rm
Let $A$ be a subset of the parameters set $E$. A pair $(f,A)$ is called a soft set with a central set $A$,
where $f$ is a mapping given by $f : E\rightarrow {\cal P}(U)$. For simply, we call $(f,A)$ a central soft set.

For two central soft sets $(f, A)$ and $(g,B)$, we say $(f, A)=(g,B)$, if $f=g$ and $A=B$.
}
\end{definition}

In fact, a central soft set $(f,A)$ over $U$ gives a complete parametrization of the universe set $U$ by the mapping $f$.
The concept presented here is different from previous ones (see \cite{first,topology}).
The central parameter set in a central soft set is to illustrate particularities which exists in information
given by a soft set.
These particularities can have a variety of meanings with different backgrounds of problems.
For example, in the instance of choosing houses,
let $(f,A)$ be a central soft set that presents information about the scoring given by Mr. X.
We can assume that $A$ is a parameters set related with his field of expertise.
To take another instance, teachers in a school who are good at teaching certain subjects
will be chosen by some students. Let $E$ be a set of school subjects,
and $U$ consists of all teachers in this school.
For each participating student $X_i$, we assume that $A_i$ is the set of his excellent courses.
A mapping $f_i: E\rightarrow {\cal P}(U)$ denotes the evaluation result by $X_i$.

The function of central sets is mainly reflected in superiority of operations between soft sets,
which will be shown in the next definitions.

\begin{definition}\label{definition:6}
{\rm The union of two central soft sets $(f, A)$ and $(g, B)$ over a common universe $U$ is a central soft set $(h, C)$,
where $C= A \cup B$ and for all $e \in E$,
$$h(e)=\left\{
          \begin {array}{ll}
          f(e),&{\mbox{if}} \ \ e\in A-B;\\
          g(e),&{\mbox{if}} \ \ e\in B-A;\\
          f(e)\cup g(e),&{\mbox{otherwise}}.
         \end{array}
        \right.$$
We write $(f,A)\sqcup (g,B)=(h,C)$.}
\end{definition}

This operation between central soft sets is given from the view of information synthesis.
Relationship between central soft sets will also be better represented by these operations defined here.

\begin{example}\label{example:5}
{\rm
Suppose that $U= \{h_1, h_2, h_3, h_4, h_5\}$ represents a set of houses.
Let $E =\{$\Rmnum{1}, \Rmnum{2}, \Rmnum{3}, \Rmnum{4}, \Rmnum{5}, \Rmnum{6}, \Rmnum{7}, \Rmnum{8}$\}$,
where roman numerals \Rmnum{1} to \Rmnum{8} represent some attributes of houses respectively such as
reasonable design space, green environment, excellent property management, convenient transportation.
Consider two central soft sets $(f,A)$ and $(g,B)$ defined over $U$,
where $A = \{\rm{\Rmnum{4},\Rmnum{5}}\}$ and $B=\{\rm{\Rmnum{1},\Rmnum{4}}\}$ are two central sets
which are precisely the areas of expertise of Mr. X and Mr. Y in choosing houses respectively,
the mappings $f$ and $g$ are defined as follows:\\
$f(\mbox{\Rmnum{1}})= \{h_2,h_3,h_4,h_5\}, \ \ \ \ f(\mbox{\Rmnum{2}})=\{h_2,h_5\}$,\\
$ f(\mbox{\Rmnum{3}})=\{h_2,h_3\}, \ \ \ \ \ \ \ \ \ \ f(\mbox{\Rmnum{4}})= \{h_1,h_4,h_5\}$,\\
$f(\mbox{\Rmnum{5}})=  \{h_1,h_4\}, \ \ \ \ \ \ \ \ \ \ \ f(\mbox{\Rmnum{6}})=\{h_1,h_5\}$,\\
$ f(\mbox{\Rmnum{7}})=\{h_2,h_5\}, \ \ \ \ \ \ \ \ \ f(\mbox{\Rmnum{8}})= \{h_3,h_4,h_5\}$;\\
and \\
$g(\mbox{\Rmnum{1}})=\{h_2,h_3,h_4\},  \ \ \ \ \ \ \ \ \ g(\mbox{\Rmnum{2}})=\{h_1,h_5\}$,\\
$ g(\mbox{\Rmnum{3}})=\{h_3,h_4\}, \ \ \ \ \ \ \ \ \ \ \ g(\mbox{\Rmnum{4}})= \{h_2,h_4,h_5\}$,\\
$g(\mbox{\Rmnum{5}})= \{h_2,h_4\},\ \ \ \ \ \ \ \ \ \ \ \ g(\mbox{\Rmnum{6}})=\{h_4,h_5\}$,\\
$ g(\mbox{\Rmnum{7}})=\{h_2,h_4\},\ \ \ \ \ \ \ \ \ \  g(\mbox{\Rmnum{8}})= \{h_1,h_4,h_5\}$.

By definitions of union and intersection on central soft sets,
we have $(f,A)\sqcup (g,B)=(h, \{\rm{\Rmnum{1},\Rmnum{4},\Rmnum{5}}\})$, where $h$ is defined as:\\
$h(\mbox{\Rmnum{1}})= \{h_2,h_3,h_4\}, \ \ \ \ \ \ \ \ \ \ h(\mbox{\Rmnum{2}})=\{h_1,h_2,h_5\}$,\\
$h(\mbox{\Rmnum{3}})=\{h_2,h_3,h_4\}, \ \ \ \ \ \ \ \ h(\mbox{\Rmnum{4}})= \{h_1,h_2,h_4,h_5\}$,\\
$h(\mbox{\Rmnum{5}})= \{h_1,h_4\}, \ \ \ \ \ \ \ \ \ \ \ \ \ \ h(\mbox{\Rmnum{6}})=\{h_1,h_4,h_5\}$,\\
$h(\mbox{\Rmnum{7}})=\{h_2,h_4,h_5\},  \ \ \ \ \ \ \ \ h(\mbox{\Rmnum{8}})= \{h_1,h_3,h_4,h_5\}$.

If we take a new central soft sets $(f,C)$, where $C=\{ \rm{\Rmnum{3}, \Rmnum{4}}\}$.
Then $(f,C)\sqcup (g,B)=(j, \{\rm{\Rmnum{1}, \Rmnum{3},\Rmnum{4}}\})$, where $j$ is defined as:\\
$j(\mbox{\Rmnum{1}})= \{h_2,h_3,h_4\}, \ \ \ \ \ \ \ \ \ \ \ j(\mbox{\Rmnum{2}})=\{h_1,h_2,h_5\}$,\\
$j(\mbox{\Rmnum{3}})=\{h_2,h_3\}, \ \ \ \ \ \ \ \ \ \ \ \ \ \ j(\mbox{\Rmnum{4}})= \{h_1,h_2,h_4,h_5\}$,\\
$j(\mbox{\Rmnum{5}})= \{h_1,h_2,h_4\}, \ \ \ \ \ \ \ \ \ \ \ j(\mbox{\Rmnum{6}})=\{h_1,h_4,h_5\}$,\\
$j(\mbox{\Rmnum{7}})=\{h_2,h_4,h_5\},  \ \ \ \ \ \ \ \ j(\mbox{\Rmnum{8}})= \{h_1,h_3,h_4,h_5\}$.

Since we choose two different central sets for the mapping $f$, two different central soft sets
$(h, \{\rm{\Rmnum{1},\Rmnum{4},\Rmnum{5}}\})$ and $(j, \{\rm{\Rmnum{1}, \Rmnum{3},\Rmnum{4}}\})$ are obtained.
It demonstrates that central sets play an important role as mappings in the operation of union.

}
\end{example}

\begin{definition}\label{definition:9}
{\rm The intersection of two central soft sets $(f, A)$ and $(g, B)$ over a common universe $U$ is a central soft set $(h, C)$,
where $C= A \cap B$, and for all $e \in E$,
$$h(e)=\left\{
          \begin {array}{ll}
          g(e),&{\mbox{if}} \ \ e\in A-B;\\
          f(e),&{\mbox{if}} \ \ e\in B-A;\\
          f(e)\cap g(e),&{\mbox{otherwise}}.
         \end{array}
        \right.$$
We write $(f,A)\sqcap (g,B)=(h,C)$.}
\end{definition}

Clearly, $(f,A)\sqcup (f,B)=(f,A\cup B)$ and $(f,A)\sqcap (f,B)=(f,A\cap B)$.

This definition may seem strange, however, actually it gives
the maximum central soft set which is contained in two original central soft sets.
It will be shown in the following conclusions.

\begin{example}\label{example:3}
{\rm As it shown in Example \ref{example:5}, for the intersection operation, we have
$(f,A)\sqcap (g,B)=(k, \{\rm{\Rmnum{4}}\})$, where $k$ is defined as:\\
$k(\mbox{\Rmnum{1}})= \{h_2,h_3,h_4,h_5\},  \ \ \ \ \ \ \ \ k(\mbox{\Rmnum{2}})=\{h_5\}$,\\
$ k(\mbox{\Rmnum{3}})=\{h_3\}, \ \ \ \ \ \ \ \ \ \ \ \ \ \ \ \ \ \ k(\mbox{\Rmnum{4}})=\{h_4,h_5\}$,\\
$k(\mbox{\Rmnum{5}})= \{h_2,h_4\}, \ \ \ \ \ \ \ \ \ \ \ \ \ \ \ k(\mbox{\Rmnum{6}})=\{h_5\}$,\\
$k(\mbox{\Rmnum{7}})=\{h_2\}, \ \ \ \ \ \ \ \ \ \ \ \ \ \ \ \ \ k(\mbox{\Rmnum{8}})=\{h_4,h_5\}$.

}
\end{example}

\begin{remark}
{\rm
In \cite{topology}, the concept of soft sets is defined as follows:
a soft set $F_A$ on the universe $U$ is defined by the set of ordered pairs
$F_A = \{(x, f_A(x)): x \in E, f_A(x)\in {\cal P}(U)\}$,
where $f_A : E \rightarrow {\cal P}(U)$ such that $f_A(x)= \emptyset$ if and only if $x \not\in A$.
Clearly there exists a correspondence between a central soft set $(f_A,A)$ and a soft set $F_A$.

Here we assume that $F_A$ and $G_B$ are two soft sets defined as above.
By selecting different central sets for mappings, we will investigate these two operations on central soft sets.

(1) We choose two central sets $A$ and $B$ for the mappings $f_A$ and $g_B$ respectively.
By Definition \ref{definition:6},
the union of two central soft sets $(f_A,A)$ and $(g_B,B)$ is a new central soft set $(h_C,C)$,
i.e., $(f_A,A)\sqcup (g_B,B)=(h_C,C)$,
where $C= A \cup B$ and for all $x \in E$,
$$h_C(x)=\left\{
          \begin {array}{ll}
          f_A(x),&{\mbox{if}} \ \ x\in A-B;\\
          g_B(x),&{\mbox{if}} \ \ x\in B-A;\\
          f_A(x)\cup g_B(x), &{\mbox{otherwise}}.
         \end{array}
        \right.$$
For all $x\not\in C$, we have $h_C(x)=\emptyset$.
Accordingly, for the mapping $h_C$ a soft set $H_C= \{(x, h_C(x)): x \in E \}$ is obtained.
Then, by the operation defined on central soft sets, actually
we define an operation $\widetilde{\sqcup}$ between soft sets $F_A$ and $G_B$, i.e., $F_A \widetilde{\sqcup} G_B=H_C$.
This definition is consistent with the operation union given in \cite{soft}.
We use a diagram to represent this relationship:
\begin{figure}[htbp]
\xymatrix@C=0.4cm@R=0.8cm 
{{\mbox{central soft sets and operations:}}&(f_A,A)\ar@<1ex>[d] &\sqcup &(g_B,B)\ar@<1ex>[d] &=(h_C,C)\ar@<1ex>[d] \\
{\mbox{soft sets and operations:}}&F_A \ar[u] &\widetilde{\sqcup} &G_B  \ar[u]
      & = H_C  \ar[u] }
\end{figure}

(2) Let $E$ be the central set for all soft sets.

The union of two central soft sets $(f_A,E)$ and
$(g_B,E)$ is a central soft set $(l,E)$, where $l$ is defined as:
\begin{center}
$l(x)= f_A(x)\cup g_B(x)$, for all $x\in E$.
\end{center}
Clearly $f_A(x)\cup g_B(x)=\emptyset$ for all $x\not\in C$, if $C=A\cup B$.
Then a soft set $L_C= \{(x, l(x)): x \in E \}$ which corresponds to $l$ is obtained.
By the operation union of central soft sets, it defines a natural union operation $\cup$ between soft sets:
\begin{center}
$F_A\cup G_B=L_C$.
\end{center}

For the intersection of two central soft sets $(f_A,E)$ and $(g_B,E)$,
let $(j,E)=(f_A,E) \sqcap (g_B,E)$, where $j$ is defined as:
\begin{center}
$j(x)= f_A(x)\cap g_B(x)$, for all $x\in E$.
\end{center}
The mapping $j$ corresponds to a soft set $J_D = \{(x, j(x)): x \in E \}$, where $D=\{x\in E: j(x)\not=\emptyset\}$.
Then the operation intersection of central soft sets just defines a natural intersection operation $\cap$ between soft sets:
\begin{center}
$F_A\cap G_B=J_D$.
\end{center}

(3)
We choose two central sets $E-A$ and $E-B$ for the mappings $f_A$ and $g_B$ respectively.

Assume that $(f_A,E-A)\sqcup (g_B,E-B)=(i,E-(A \cap B))$. For all $x \in E$,
$$i(x)=\left\{
          \begin {array}{ll}
          g_B(x),&{\mbox{if}} \ \ x\in A-B;\\
          f_A(x),&{\mbox{if}} \ \ x\in B-A;\\
          f_A(x)\cup g_B(x), &{\mbox{otherwise}}.
         \end{array}
        \right.$$
Let $C=A \cup B$. For all $x\not\in C$, we have $i(x)=\emptyset$.
Accordingly, for the mapping $i$ a soft set $I_C= \{(x, i(x)): x \in E \}$ is obtained.
Then
we have a new operation $\breve{\sqcup}$ between soft sets $F_A$ and $G_B$, i.e., $F_A \breve{\sqcup} G_B=I_C$.

Assume that $(f_A,E-A)\sqcap (g_B,E-B)=(k,E- (A \cup B))$. For all $x \in E$,
$$k(x)=\left\{
          \begin {array}{ll}
          f_A(x),&{\mbox{if}} \ \ x\in A-B;\\
          g_B(x),&{\mbox{if}} \ \ x\in B-A;\\
          f_A(x)\cap g_B(x), &{\mbox{otherwise}}.
         \end{array}
        \right.$$
Accordingly, for the mapping $k$ a soft set $K_D= \{(x, k(x)): x \in E \}$ is obtained, where $D=\{x\in E: k(x)\not=\emptyset\}$.
Then the operation union of central soft sets gives a new operation
\begin{center}
$F_A \hat{\sqcap} G_B=K_D$.
\end{center}
which is similar to
the extended intersection $\sqcap_\varepsilon$ between soft sets(see \cite{operation}).


As discussed above, central sets and mappings play same important role in operations between central soft sets.
In one sense,
central soft sets and operations defined above can be looked as a generalization of classic soft sets and some related operations.
}

\end{remark}

\begin{definition}\label{definition:3}
{\rm For two central soft sets $(f,A)$ and $(g,B)$ over a common universe $U$, we write $(f,A)\sqsubseteq (g,B)$,
and call it a central soft information order if $(f,A)\sqcup(g,B)=(g,B)$.
}
\end{definition}

The following equivalent forms of this ordering defined between central soft sets can be obtained immediately
by definitions above.

\begin{proposition}\label{proposition:2}
{\rm For two central soft sets $(f,A)$ and $(g,B)$ over a common universe $U$, $(f,A)\sqsubseteq (g,B)$,
if and only if any of the following conditions are true

(i) $A \subseteq B$ and for all $e\not\in B-A$, $f(e)\subseteq g(e)$;

(ii) $(f,A)\sqcap (g,B)=(f,A)$.

}
\end{proposition}

By Proposition \ref{proposition:2}, we can directly show that the central soft information order is antisymmetric and transitive.
Therefore, this ordering is a partial order.

\begin{example}\label{example:6}
{\rm As it shown in Example \ref{example:5},
let $C =\{\rm{\Rmnum{1},\Rmnum{4},\Rmnum{6}}\}$ be a central set,
and the mapping $m: E\rightarrow {\cal P}(U)$ is defined as follows:\\
$m({\rm{\Rmnum{1}}})= \{h_2,h_3,h_4\},\ \ \ \ \ \ \ \ \ m(\mbox{\Rmnum{2}})=\{h_1,h_5\}$,\\
$ m(\mbox{\Rmnum{3}})=\{h_3,h_4\},\ \ \ \ \ \ \ \ \ \ \ m(\mbox{\Rmnum{4}})=\{h_2,h_4,h_5\}$,\\
$m(\mbox{\Rmnum{5}})= \{h_2,h_4\},\ \ \ \ \ \ \ \ \ \ \ \  m(\mbox{\Rmnum{6}})=\{h_1\}$,\\
$m(\mbox{\Rmnum{7}})=\{h_2,h_4\},\ \ \ \ \ \ \ \ \ \  m(\mbox{\Rmnum{8}})=\{h_1,h_4,h_5\}$.\\
We have $(g, B)\sqsubseteq (m,C)$.

By the formula $(g,B)\sqsubseteq (m,C)$ showed in this example,
we can intuitively understand the meaning of central soft information order from experience.
Suppose that $(m,C)$ means a result of evaluation of Mr. Z for choosing houses.
The relation $B \subseteq C$ says that Mr. Z has a wider professional knowledge than Mr. X.
Meanwhile, for each parameter $e$ not in the set $C-B$ which is an exclusive advantage region of Mr. Z,
$m(e)$ is stronger than $g(e)$ when we combine this data.

}
\end{example}

\begin{definition}\label{definition:10}
{\rm The complement of a central soft set $(f, A)$ is denoted by $(f, A)^c=(f^c, A^c)$,
where $A^c=E-A$, $f^c(e)=U-f(e)$ for all $e\in E$.}
\end{definition}

\begin{definition}\label{definition:2}
{\rm For two central soft sets $(f, A)$ and $(g, B)$ over a common universe $U$,
$(f, A)-(g, B)$ is defined to be a central soft set $(f, A)\sqcap(g, B)^c$.

}
\end{definition}

In fact, let $(f, A)\sqcap(g, B)^c=(h, C)$, then $C=A-B$ and $$h(e)=\left\{
          \begin {array}{ll}
          g^c(e),&{\mbox{if}} \ \ e\in A\cap B;\\
          f(e),&{\mbox{if}} \ \ e\in (A\cup B)^c;\\
          f(e)\cap G^c(e),&{\mbox{otherwise}}.
         \end{array}
        \right.$$
For $e\not\in A\cap B$, $h(e)\subseteq f(e)$. According to Definition \ref{definition:3},
we have $(f, A)-(g, B)\sqsubseteq(f,A)$.
It accords with the general characteristics of this operation.

\begin{example}\label{example:2}
{\rm As it shown in Example \ref{example:6},
$(m,C)-(g, B)=(n,\{\mbox{\Rmnum{6}}\})$,
where the mapping $n: E\rightarrow {\cal P}(U)$ is defined as follows:\\
$n({\rm{\Rmnum{1}}})= \emptyset,\ \ \ \ \ \ \ \ \ \ \ \ \ \ \ \ \ \ \ \ \ \ n(\mbox{\Rmnum{2}})=\{h_1,h_5\}$,\\
$n(\mbox{\Rmnum{3}})=\{h_3,h_4\},\ \ \ \ \ \ \ \ \ \ \ n(\mbox{\Rmnum{4}})=\emptyset$,\\
$n(\mbox{\Rmnum{5}})= \{h_2,h_4\},\ \ \ \ \ \ \ \ \ \ \ \  n(\mbox{\Rmnum{6}})=\{h_2,h_3,h_4,h_5\}$,\\
$n(\mbox{\Rmnum{7}})=\{h_2,h_4\},\ \ \ \ \ \ \ \ \ \  n(\mbox{\Rmnum{8}})=\{h_1,h_4,h_5\}$.\\

}
\end{example}

\section{Properties of operations on central soft sets}
In this part we study the properties of operations on central soft sets which are over a same universe set.

\begin{theorem}\label{theorem:1}
{\rm Let $(f, A)$ and $(g, B)$ be two central soft sets over a same universe $U$. Then

1. $[(f,A)\sqcap (g,B)]^c=(f,A)^c\sqcup (g,B)^c$;

2. $[(f,A)\sqcup (g,B)]^c=(f,A)^c\sqcap (g,B)^c$.
}
\end{theorem}
\noindent {\bf Proof.}
Clearly the central sets in both sides of the first equation are $(A\cap B)^c$.
We write $(f,A)\sqcap (g,B)=(h, A\cap B)$ and $(f,A)^c\sqcup (g,B)^c=(k, A^c\cup B^c)$.

Take any $e\in E$. If $e\in A^c-B^c$, i.e., $e\in B-A$, we have $h^c(e)=U-h(e)=U-f(e)=f^c(e)=k(e)$.
If $e\in B^c-A^c$, then $h^c(e)=g^c(e)=k(e)$ similarly.
Otherwise $e\not\in (A-B)\cup (B-A)$, then $h^c(e)=U-h(e)=U-(f(e)\cap g(e))=f^c(e)\cup g^c(e)=k(e)$.
In summary, we have $[(f,A)\sqcap (g,B)]^c=(f,A)^c\sqcup (g,B)^c$.

According to the proof above, we can also show the second equation directly.
\qed

The associative laws are also true for two operations union and intersection defined here.
\begin{theorem}\label{theorem:2}
{\rm Let $(f, A)$, $(g, B)$ and $(h, C)$ be central soft sets over a same universe set $U$. Then

1. $(f,A)\sqcap[(g,B)\sqcap(h, C)]=[(f,A)\sqcap(g,B)]\sqcap(h, C)$;

2. $(f,A)\sqcup[(g,B)\sqcup(h, C)]=[(f,A)\sqcup(g,B)]\sqcup(h, C)$.
}
\end{theorem}
\noindent {\bf Proof.}
Let write\\
$(g,B)\sqcap (h,C)=(j, B\cap C)$;\\
$(f,A)\sqcap (g,B)=(k, A\cap B)$;\\
$(f,A)\sqcap (j, B\cap C)=(l,A\cap B\cap C)$;\\
$(k, A\cap B) \sqcap(h, C)=(m,A\cap B\cap C)$.\\
We will show that $l(e)=m(e)$ for all $e\in E$ in order to prove the associative law of the operation intersection.

(1) If $e\in (B\cap C)-A$, then $l(e)=f(e)$. Since $e\in C-(A\cap B)$, $m(e)=k(e)=f(e).$ Thus $m(e)=l(e)$.

(2) If $e\in A-(B \cap C)$, it can be divided into three conditions:\\
(\rmnum{1}) If $e\in A$, $e\not\in B$ and $e\not\in C$, then $l(e)=j(e)=g(e)\cap h(e)$ and $m(e)=k(e)\cap h(e)=g(e)\cap h(e)$.\\
(\rmnum{2}) If $e\in A$, $e\not\in B$ and $e\in C$, then $l(e)=j(e)=g(e)$, $m(e)=k(e)=g(e)$.\\
(\rmnum{3}) If $e\in A$, $e\not\in C$ and $e\in B$, then $l(e)=j(e)=h(e)$, $m(e)=h(e)$.

(3) If $e\not\in A-(B \cap C)$ and $e\not\in (B \cap C)-A$, then $l(e)=m(e)=f(e)\cap g(e)\cap h(e)$.\\
Thus we obtain that $l(e)=m(e)$ for all $e\in E$.

The proof of the next equation is omitted here.
\qed

\begin{theorem}\label{theorem:3}
{\rm Let $(f, A)$, $(g, B)$ and $(h, C)$ be central soft sets over a same universe $U$. Then

1. $(f,A)\sqcap[(g,B)\sqcup(h, C)]=[(f,A)\sqcap(g,B)]\sqcup [(f,A)\sqcap(h, C)]$;

2. $(f,A)\sqcup[(g,B)\sqcap(h, C)]=[(f,A)\sqcup(g,B)]\sqcap [(f,A)\sqcup(h, C)]$.
}
\end{theorem}
\noindent {\bf Proof.} Here we only prove the first equation.
We write\\
$(f,A)\sqcap (g,B)=(i, A\cap B)$;\\
$(f,A)\sqcap (h,C)=(j, A\cap C)$;\\
$(g,B)\sqcup (h,C)=(k, B\cup C)$;\\
$(f,A)\sqcap (k, B\cup C)=(l,A\cap(B\cup C))$;\\
$(i, A\cap B) \sqcup(j, A\cap C)=(m,(A\cap B)\cup (A\cap C))$.\\
Then we need to show that $l(e)=m(e)$ for all $e\in E$. It can be divided into the following several kinds:

(1) If $e\in A-(B \cup C)$, then $l(e)=k(e)=g(e)\cup h(e)$, $i(e)=g(e)$ and $j(e)=h(e)$.
Since $e\not\in A\cap B$ and $e\not\in A\cap C$, we have
$m(e)=i(e)\cup j(e)=g(e)\cup h(e)=l(e)$.

(2) If $e\in (B\cup C)-A$, then $l(e)=f(e)$. Since $e\not\in A\cap B$ and $e\not\in A\cap C$, we have $m(e)=i(e)\cup j(e)$.
The set $(B\cup C)-A$ can be divided into three mutually disjoint parts $S_1=(B-A)-(B\cap C)$, $S_2=(C-A)-(B\cap C)$
and $S_3=(B\cap C)-A$.\\
(\rmnum{1}) If $e\in S_1$, $i(e)=f(e)$ and $j(e)=f(e)\cap h(e)$. Then $m(e)=i(e)\cup j(e)=f(e)$.\\
(\rmnum{2}) If $e\in S_2$, $j(e)=f(e)$ and $i(e)=f(e)\cap g(e)$. Then $m(e)=i(e)\cup j(e)=f(e)$.\\
(\rmnum{3}) If $e\in S_3$, $i(e)=j(e)=f(e)$. Then $m(e)=i(e)\cup j(e)=f(e)$.\\
Thus we obtain that $l(e)=f(e)=m(e)$ for all $e\in (B\cup C)-A$.

(3) Otherwise $e$ is in the complementary set of $[(B\cup C)-A]\cup [A-(B\cup C)]$, then $l(e)=f(e)\cap k(e)$.
The complementary set of $[(B\cup C)-A]\cup [A-(B\cup C)]$ can be divided into four mutually disjoint parts
\begin{center}
$S_4=(A\cap B)-(B\cap C)$, $S_5=(A\cap C)-(B\cap C)$, \\
$S_6=A\cap B\cap C$, and $S_7=E-(A\cup B\cup C)$.
\end{center}
(\rmnum{1}) If $e\in S_4$, then $k(e)=g(e)$ and $m(e)=i(e)=f(e)\cap g(e)=f(e)\cap k(e)=l(e)$.\\
(\rmnum{2}) If $e\in S_5$, then $k(e)=h(e)$ and $m(e)=j(e)=f(e)\cap h(e)=l(e)$.\\
(\rmnum{3}) If $e\in S_6$ or $e\in S_7$, then
$k(e)=g(e)\cup h(e)$ and $$m(e)=i(e)\cup j(e)=[f(e)\cap g(e)]\cup [f(e)\cap h(e)]=f(e)\cap[g(e)\cup h(e)]
=f(e)\cap k(e)=l(e).$$
Thus $l(e)=m(e)$ for all $e\in E-[(B\cup C)-A]- [A-(B\cup C)]$.

By the proof above we have $l(e)=m(e)$ for all $e\in E$. So the first equation has been shown.
By the same way the next distributive property of union with respect to intersection can be shown.
\qed

Next we get a natural property which similar as properties of operations of addition and subtraction of numbers.

\begin{theorem}\label{theorem:4}
{\rm Let $(f, A)$, $(g, B)$ and $(h, C)$ be central soft sets over a same universe $U$.
Then
\begin{center}
$(f, A)-(g, B)-(h, C)=(f, A)-[(g, B)\sqcup(h, C)]$.
\end{center}
}
\end{theorem}
\noindent {\bf Proof.}
According to the definition of the operation subtraction $-$ and Theorem \ref{theorem:1}, we have
\begin{eqnarray*}
\begin{array}{lll}
(f, A)-(g, B)-(h, C)
&=& (f, A)\sqcap(g, B)^c\sqcap(h, C)^c\\
&=& (f, A)\sqcap[(g, B)\sqcup(h, C)]^c\\
&=& (f, A)-[(g, B)\sqcup(h, C)].
\end{array}
\end{eqnarray*}
\qed

\begin{theorem}\label{theorem:5}
{\rm Let $\{(f_i, A_i): i\in I\}$ be a set of central soft sets over a same universe $U$. Then
\begin{center}
$\mathop\bigsqcup\limits_{i\in I}(f_i, A_i) = (h, \mathop\bigcup\limits_{i\in I} A_i),$
\end{center}
where $h: E \rightarrow {\cal P}(U)$ is defined as follows:

$\forall e\in E$, let $I_e=\{i\in I: e\in A_i\}$, we have
$$h(e)=\left\{
          \begin {array}{ll}
          \mathop\bigcup\limits_{i\in I_e} f_i(e),&{\mbox{if}} \ \ I_e\not=\emptyset;\\

          \mathop\bigcup\limits_{i\in I} f_i(e),&{\mbox{otherwise}}.
         \end{array}
        \right.$$
 }
\end{theorem}
\noindent {\bf Proof.}
First we show that $(h,\mathop\bigcup\limits_{i\in I} A_i)$ is an upper bound of $\{(f_i, A_i): i\in I\}$
by Proposition \ref{proposition:2}.
Let $j\in I$ and $e\not\in\mathop\bigcup\limits_{i\in I} A_i-A_j$.
If $I_e\not=\emptyset$, then $j\in I_e$. We have $h(e)=\mathop\bigcup\limits_{i\in I_e} f_i(e)\supseteq f_j(e)$.
By the definition of $h$, we show that $f_j(e)\subseteq h(e)$ for $e\not\in\mathop\bigcup\limits_{i\in I} A_i-A_j$.
Then $(f_j, A_j)\sqsubseteq (h,\mathop\bigcup\limits_{i\in I} A_i)$.

Suppose that $(g,B)$ is another upper bound of the set $\{(f_i, A_i): i\in I\}$.
Clearly we have $\mathop\bigcup\limits_{i\in I} A_i\subseteq B$.
In order to show that $(h, \mathop\bigcup\limits_{i\in I} A_i)\sqsubseteq (g,B)$,
we only need to prove $h(e)\subseteq g(e)$ for $e\not\in B-\mathop\bigcup\limits_{i\in I} A_i$.
In fact, if $e\not\in B$, then $e\not\in B-A_i$ for all $i\in I$. So $f_i(e)\subseteq g(e)$.
Thus $h(e)\subseteq \mathop\bigcup\limits_{i\in I} f_i(e)\subseteq g(e)$.
If $e\in \mathop\bigcup\limits_{i\in I} A_i$, then $e\not\in B-A_j$ for all $j\in I_e$.
We obtain that $f_j(e)\subseteq g(e)$ for all $j\in I_e$.
Thus $h(e)=\mathop\bigcup\limits_{j\in I_e} f_j(e)\subseteq g(e)$.
So $(h, \mathop\bigcup\limits_{i\in I} A_i)\sqsubseteq (g,B)$.
By the definition of supremum and what we have proved above, it follows that
$\mathop\bigsqcup\limits_{i\in I}(f_i, A_i) = (h, \mathop\bigcup\limits_{i\in I} A_i).$
\qed

\begin{theorem}\label{theorem:7}
{\rm Let $\{(f_i, A_i): i\in I\}$ be a set of central soft sets over a same universe $U$. Then
\begin{center}
$(f,A)\sqcap \mathop\bigsqcup\limits_{i\in I}(f_i, A_i) = \mathop\bigsqcup\limits_{i\in I} [(f,A)\sqcap (f_i, A_i)],$
\end{center}
 }
\end{theorem}
\noindent {\bf Proof.}
Let
\begin{center}
$\bigsqcup\limits_{i\in I}(f_i, A_i)=(g, \mathop\bigcup\limits_{i\in I} A_i),$\\
$(f,A)\sqcap (g, \mathop\bigcup\limits_{i\in I} A_i)=(k, A\cap (\bigcup\limits_{i\in I} A_i))$,\\
$(f,A)\sqcap (f_i,A_i)=(h_i, A\cap A_i),$\\
$\bigsqcup\limits_{i\in I}(h_i, A\cap A_i)=(h, A\cap (\bigcup\limits_{i\in I} A_i)).$
\end{center}
Following we need to show that $k=h$.

If $e\in A- \bigcup\limits_{i\in I} A_i$, then $k(e)=g(e)$.
Since $\{i\in I: e\in A\cap A_i\}=\emptyset$ and $\{i\in I: e\in A_i\}=\emptyset$, by Theorem \ref{theorem:5} we have
$h(e)=\mathop\bigcup\limits_{i\in I} h_i(e)=\mathop\bigcup\limits_{i\in I} f_i(e)=g(e)$.

If $e\in (\bigcup\limits_{i\in I} A_i)-A$, then $k(e)=f(e)$.
Since $\{i\in I: e\in A\cap A_i\}=\emptyset$, by Theorem \ref{theorem:5} we have
$h(e)=\mathop\bigcup\limits_{i\in I} h_i(e)$.
For $i\in I$ such that $e\in A_i-A$, $h_i(e)=f(e)$. While, for $i\in I$ such that $e\not\in A_i-A$, $h_i(e)=f(e)\cap f_i(e)$.
So $h(e)=\mathop\bigcup\limits_{i\in I} h_i(e)=f(e)$.

Otherwise, for $e\not\in A- \bigcup\limits_{i\in I} A_i$ and $e\not\in (\bigcup\limits_{i\in I} A_i)-A$, we have\\
\begin{center}
$k(e)=f(e)\cap g(e)=f(e)\cap \mathop\bigcup\limits_{i\in I} f_i(e)=\mathop\bigcup\limits_{i\in I} [f(e)\cap f_i(e)]
=\mathop\bigcup\limits_{i\in I} h_i(e)=h(e).$
\end{center}
In any case we obtain that $k(e)=h(e)$.
\qed

\begin{definition}\label{definition:11}
{\rm Let $(f,A)$ be a central soft set over $U$ and $B\subseteq A$.
The projection of $(f,A)$ over $B$, written $(f,A)^{\downarrow B}$, is defined to be a new central soft
set $(g,B)$ such that $g(e) = f(e)$ for all $e\in E$.}
\end{definition}

The purpose of the projection operation is to constrain these central sets of central soft sets.
While, mappings of central soft sets will not be changed.

\begin{example}\label{example:1}
{\rm
As it shown in Example \ref{example:5},
if $S=\{\rm{\Rmnum{1},\Rmnum{5}}\}$, then $S \cap A=\{\Rmnum{5}\}, S \cap B=\{\Rmnum{1}\}$.
Thus $(f,A)^{\downarrow {S \cap A}}\sqcup (g,B)^{\downarrow {S \cap B}}$ gives a new central soft set $(k,S)$
defined as:\\
$k(\mbox{\Rmnum{1}})= \{h_2,h_3,h_4\}, \ \ \ \ \ \ \ \ \ \ k(\mbox{\Rmnum{2}})=\{h_1,h_2,h_5\}$,\\
$k(\mbox{\Rmnum{3}})=\{h_2,h_3,h_4\}, \ \ \ \ \ \ \ \ k(\mbox{\Rmnum{4}})= \{h_1,h_2,h_4,h_5\}$,\\
$k(\mbox{\Rmnum{5}})= \{h_1,h_4\}, \ \ \ \ \ \ \ \ \ \ \ \ \ \ k(\mbox{\Rmnum{6}})=\{h_1,h_4,h_5\}$,\\
$k(\mbox{\Rmnum{7}})=\{h_2,h_4,h_5\},  \ \ \ \ \ \ \ \ k(\mbox{\Rmnum{8}})= \{h_1,h_3,h_4,h_5\}$.

According to the conclusion of Example \ref{example:5}, we have $k=h$. Then
$$[(f,A) \sqcup (g,B)]^{ \downarrow S}=(f,A)^{\downarrow {S \cap A}}\sqcup (g,B)^{\downarrow {S \cap B}}.$$
}
\end{example}

In fact, we can show the following general conclusion.
\begin{proposition}\label{proposition:1}
{\rm  For two central soft sets $(f,A)$ and $(g,B)$,
$$[(f,A) \sqcup (g,B)]^{ \downarrow S} \sqsubseteq (f,A)^{\downarrow {S \cap A}}\sqcup (g,B)^{\downarrow {S \cap B}},$$
if $S\subseteq A\cup B$. Especially,
$$[(f,A) \sqcup (g,B)]^{ \downarrow S}=(f,A)^{\downarrow {S \cap A}}\sqcup (g,B)^{\downarrow {S \cap B}},$$
if $S=(A\cup B)-D$, where $D\subseteq A\cap B$.
 }
\end{proposition}
\noindent {\bf Proof.} Let
$$(f,A) \sqcup (g,B)=(h, A\cup B)$$
and
$$(f,A)^{\downarrow {S \cap A}}\sqcup (g,B)^{\downarrow {S \cap B}}=(h^{'}, S\cap (A\cup B)).$$
For all $e\in E$, we have
$$h^{'}(e)=\left\{
          \begin {array}{ll}
          f(e),&{\mbox{if}} \ \ e\in (S\cap A)-(S\cap B);\\
          g(e),&{\mbox{if}} \ \ e\in (S\cap B)-(S\cap A);\\
          f(e)\cup g(e),&{\mbox{otherwise}} .
          \end{array}
          \right.
          $$

Clearly $(S\cap A)-(S\cap B) \subseteq A-B$ and $(S\cap B)-(S\cap A) \subseteq B-A$.
Then we have $h(e)\subseteq h^{'} (e)$ for all $e\in E$ by the definition of $h$.
By Proposition \ref{proposition:2}, we obtain that
$$[(f,A) \sqcup (g,B)]^{ \downarrow S} \sqsubseteq (f,A)^{\downarrow {S \cap A}}\sqcup (g,B)^{\downarrow {S \cap B}}.$$

If $S=(A\cup B)-D$ and $D\subseteq A\cap B$, then we can show that
$(S\cap A)-(S\cap B)=A-B$ and $(S\cap B)-(S\cap A)=B-A$.
By the definitions of $h$ and $h^{'}$, we obtain that
\begin{center}
\hspace{3cm}
$[(f,A) \sqcup (g,B)]^{ \downarrow S}=(f,A)^{\downarrow {S \cap A}}\sqcup (g,B)^{\downarrow {S \cap B}}.$
\qed
\end{center}

Similarly the conclusion on the intersection operation also can be obtained.

\begin{proposition}\label{proposition:3}
{\rm  For two central soft sets $(f,A)$ and $(g,B)$, if $ S\subseteq A\cap B$, then
$$(f,A)^{\downarrow S} \sqcap (g,B)^{\downarrow S} \sqsubseteq [(f,A) \sqcap (g,B)]^{\downarrow S}.$$
}
\end{proposition}

These properties of operations given in Proposition \ref{proposition:3} and Proposition \ref{proposition:1}
show some basic regulation of information synthesis (union operation and intersection operation).

\section{Evaluation systems and solutions of central soft sets}

Let $(f,A)$ be a central soft set and $E$ be a set of initial parameters. If a parameters set $D\subseteq E$ is such that\\
(1) $A\subseteq D$;\\
(2) $|D|=\max\{|G|: A\subseteq G, \bigcap\limits_{e\in G} f(e)\not=\emptyset\}$,\\
then $x\in \bigcap\limits_{e\in D} f(e)$ is called an optional solution of $(f,A)$.
Especially, if $\bigcap\limits_{e\in E} f(e)\not=\emptyset$, we call
$x\in \bigcap\limits_{e\in E} f(e)$ a perfect solution of $(f,A)$.

Let $A\subseteq E$ be a set of parameters, $\{(f_i, A_i): i\in I\}$ is a set of central soft sets.
If $A\subseteq \bigcup\limits_{i\in I} A_i$, then we call $\{(f_i, A_i): i\in I\}$ an evaluation system for the parameters set $A$.
The optional solutions of $[\bigsqcup\limits_{i\in I} (f_i, A_i)]^{\downarrow A}$
are called the optional solutions of this evaluation system.

Following we will give an algorithm for obtaining solution of a central soft set and an evaluation system.\\
{\bf An algorithm for obtaining solution of a central soft set:}
Let $U=\{o_1,o_2,\cdots,o_m\}$ be a universal set and $A$ be a subset of a parameters set $E$.

1. Give an order for the elements of $E$, and we denote it by $E=\{e_1,e_2,\cdots,e_n\}$.

2. Take a matrix $L_{n\times m}$, where the elements $l_{ij}(i=1,2,\cdots,n; j=1,2,\cdots,m)$ are defined as follows:
$$l_{ij}=\left\{
          \begin{array}{ll}
          1,&{\mbox{if}} \ \ o_j\in F(e_i);\\

          0,&{\mbox{otherwise}}.
         \end{array}
        \right.$$

3. Compute $\sum\limits_{i: e_i\in A} l_{ij}$ and $\sum\limits_{i} l_{ij}$ for each a fixed $j$,
and denote it by $b_j$ and $a_j$ respectively.
If the index set $J=\{j: b_j=|A|\}\not=\emptyset$, let $j^*\in J$ be a index such that
$a_{j^{*}}=\max\{a_j: b_j=|A|\}$, then $o_{j^{*}}$ is an optional solution of $(f,A)$.

\begin{example}\label{example:4}
{\rm

As the example shown in Example \ref{example:5}, we can get two $8\times 5$ matrices $L, M$
shown as follows for the two central soft sets $(f,A)$ and $(g,B)$:

\begin{equation*}                  
L =\left(                             
\begin{array}{ccccc}
0&1&1&1&1\\
0&1&0&0&1\\
0&1&1&0&0\\
1&0&0&1&1\\
1&0&0&1&0\\
1&0&0&0&1\\
0&1&0&0&1\\
0&0&1&1&1
\end{array}
\right)
\ \ \ \ \ \ \ \ \ \ \ \ \ \ \ \ \ \                             
M=\left(
\begin{array}{ccccc}                   
0&1&1&1&0\\
1&0&0&0&1\\
0&0&1&1&0\\
0&1&0&1&1\\
0&1&0&1&0\\
0&0&0&1&1\\
0&1&0&1&0\\
1&0&0&1&1
\end{array}
\right)                            
\end{equation*}                  

For the matrix $L$, we have $b_1=b_4=2$, $b_2=b _3=0, b_5=1$, and $a_1=3, a_2=4,a_3=3,a_4=4,a_5=6$.
According to the definitions shown in the above, we know that $h_4$ is the optimal solution of $(f,A)$.

For the matrix $M$, we have $b_1=0$, $b_2=b_4=2, b_3=b_5=1$, and $a_1=2, a_2=4,a_3=2,a_4=7,a_5=4$.
Then $h_4$ is also the optimal solution of $(g,B)$.

}

\end{example}

\begin{example}\label{example:7}
{\rm
As the example shown in Example \ref{example:5},
suppose that a parameter set $D=\{\rm{\Rmnum{1},\Rmnum{4},\Rmnum{5}}\}$,
then $\{(f,A), (g,B)\}$ is an evaluation system for $D$ clearly.
We have that $(f,A)\sqcup (g,B)=(h, D)$, where $h$ is defined as follows:\\
$h(\mbox{\Rmnum{1}})=\{h_2,h_3,h_4\},\ \ \ \ \ \ \ \ \ \ \ \ h(\mbox{\Rmnum{2}})=\{h_1,h_2,h_5\}$,\\
$ h(\mbox{\Rmnum{3}})=\{h_2,h_3,h_4\},\ \ \ \ \ \ \ \ \ \ h(\mbox{\Rmnum{4}})= \{h_1,h_2,h_4,h_5\}$,\\
$h(\mbox{\Rmnum{5}})=  \{h_1,h_4\}, \ \ \ \ \ \ \ \ \ \ \ \ \ \ \ h(\mbox{\Rmnum{6}})=\{h_1,h_4,h_5\}$,\\
$ h(\mbox{\Rmnum{7}})=\{h_2,h_4,h_5\}, \ \ \ \ \ \ \ \ \ h(\mbox{\Rmnum{8}})= \{h_1,h_3,h_4,h_5\}$.\\
The following matrix $N$ is corresponded to this central soft set $(h, D)$:

\begin{equation*}                  
N =\left(                             
\begin{array}{ccccc}
0&1&1&1&0\\
1&1&0&0&1\\
0&1&1&1&0\\
1&1&0&1&1\\
1&0&0&1&0\\
1&0&0&1&1\\
0&1&0&1&1\\
1&0&1&1&1
\end{array}
\right).
\end{equation*}
$h_4$ is an optimal solutions of this evaluation system.
There is no any perfect solution.
}

\end{example}

\section{Conclusions}\label{sec:con}
The concept of central soft sets is introduced in this paper.
Properties of some operations such as union, intersection, complement and projection on central soft sets are shown.
Evaluation systems and theirs optional solutions for central soft sets are proposed.
An algorithm of giving optional solutions to solve such decision making problems is presented.

\section*{Acknowledgments}
This work is supported by National Science Foundation of China (Grant No.61300153).


\begin{thebibliography}{stringX}
\leftskip=-10mm
\parskip=-3mm
\footnotesize
\bibitem{first} D. Molodtsov, Soft set theory-First results, Comput. Math. Appl. 37(1999) 19-31.

\bibitem{optimization} D. Kovkov, V. Kolbanov, D. Molodtsov, Soft sets theory-based optimization,
 Journal of Computer and Systems Sciences International 46 (2007) 872--880.

\bibitem{soft} P.K. Maji, R. Biswas, A.R. Roy, Soft set theory, Comput. Math. Appl. 45(2003) 555-562.

\bibitem{rings}
U. Acar, F. Koyuncu, B. Tanay, Soft sets and soft rings, Comput. Math. Appl. 59(2010) 3458-3463.

\bibitem{group} H. Akta\c{s}, N. \c{C}a\u{g}man, Soft sets and soft groups, Inform. Sci. 177(2007) 2726-2735.

\bibitem{semirings} F. Feng, Y.B. Jun, X.Z. Zhao, Soft semirings, Comput. Math. Appl. 56(2008) 2621-2628.

\bibitem{algebra} Y.B. Jun, Soft BCK/BCI-algebras, Comput. Math. Appl. 56(2008) 1408-1413.

\bibitem{operations} A. Sezgina, A.O. Atag\"{u}n, On operations of soft sets, Comput. Math. Appl. 61(2011) 1457-1467.

\bibitem{operation} M. Irfan Ali, F. Feng,  X.Y. Liu, W.K. Min, M. Shabir,
On some new operations in soft set theory, Comput. Math. Appl.57 (2009) 1547--1553.

\bibitem{continuous}
X.C. Guan, Y.M. Li, F. Feng, A new order relation on fuzzy soft sets and its application,
Soft Computing 17(2013) 63--70.


\bibitem{lattice}
K.Y. Qin, H. Zhao, Lattice Structures of Fuzzy Soft Sets, Lecture Notes in Computer Science 6215(2010) 126-133.

\bibitem{ideal} Y.B. Jun, C.H. Park, Applications of soft sets in ideal theory of BCK/BCI-algebras, Inform. Sci.
178(2008) 2466-2475.

\bibitem{Feng} F. Feng, C.X Li, B. Davvaz, M. Irfan Ali,
Soft sets combined with fuzzy sets and rough sets: a tentative approach, Soft Computing 14(2010) 899--911.

\bibitem{Aktas} H. Akta\c{s}, Some algebraic applications of soft sets, Applied Soft Computing 28(2015) 327--331.

 \bibitem{Babitha} K.V. Babitha, J.J. Sunil, Soft set relations and functions,
Comput. Math. Appl. 60(7)(2010) 1840-1849.

\bibitem{fuzzy} P.K. Maji, R. Biswas, A.R. Roy, Fuzzy Soft Sets, Journal of Fuzzy Mathematics 9(2001) 589--602.


\bibitem{interval}
M. Son, Interval-valued Fuzzy Soft Sets, Journal of Korean Institute of Intelligent Systems 4(2007) 557--562.

\bibitem{int} P.K. Maji, R. Biswas, A.R. Roy, Intuitionistic fuzzy soft sets, Journal of Fuzzy Mathematics 9 (2001) 677--691.

\bibitem{topology}
N. \c{C}a\u{g}man, S. Karatas, S. Enginoglu, Soft topology, Comput. Math. Appl. 62(2011) 351--358.

\bibitem{category}
O. Zahiri, Category of soft sets, Annals of the University of Craiova, Mathematics and Computer Science Series
40(2013) 154--166.

\bibitem{Prob} P. Zhu, Q.Y. Wen, Probabilistic Soft Sets, 2010 IEEE International Conference on Granular Computing
 635--638.

\bibitem{logic}
Y. Jiang, Y. Tang, Q. Chen, H. Liu, J.C Tang, Extending fuzzy soft sets with fuzzy description logics,
Knowledge-Based Systems 24(2011) 1096--1107.


\bibitem{reduction} D. Chen, E.C.C. Tsang, D.S. Yeung, X. Wang, The parametrization reduction of soft sets
and its applications, Comput. Math. Appl. 49 (2005) 757-763.

\bibitem{decision} A.R. Roy, P.K. Maji, A fuzzy soft set theoretic approach to decision making problems,
J. Comput. Appl. Math. 203 (2007) 412-418.

\bibitem{Deli} I. Deli, N. \c{C}a\u{g}man,
Intuitionistic fuzzy parameterized soft set theory and its decision making, Applied Soft Computing 28(2015) 109-113.

\bibitem{adjustable} F. Feng, Y. B. Jun, X.Y. Liu, L.F. Li,
An adjustable approach to fuzzy soft set based  decision making, J. Comput. Appl. Math. 234 (2010) 10--20.

\end{thebibliography}
\end{document}